\documentclass[10pt, prd, twocolumn, nofootinbib,preprint,superscriptaddress]{revtex4}
\pdfoutput=1
\usepackage[T1]{fontenc}
\usepackage{comment}
\usepackage{amsmath,amssymb,ulem}
\usepackage{epsfig}
\usepackage{graphicx}
\usepackage[usenames,dvipsnames]{color}
\usepackage{subfigure}
\usepackage{slashed}
\usepackage[colorlinks,citecolor=blue]{hyperref}
\usepackage{pdfpages}
\usepackage{color}

\begin{document}
\title{Gravitational waves from a dark $U(1)_D$ phase transition \\ in the light of NANOGrav 12.5 yr data}

\author{Debasish Borah}
\email{dborah@iitg.ac.in}
\affiliation{Department of Physics, Indian Institute of Technology Guwahati, Assam 781039, India}
\author{Arnab Dasgupta}
\email{arnabdasgupta@protonmail.ch}
\affiliation{Institute of Convergence Fundamental Studies , Seoul-Tech, Seoul 139-743, Korea}
\author{Sin Kyu Kang}
\email{skkang@seoultech.ac.kr}
\affiliation{School of Liberal Arts, Seoul-Tech, Seoul 139-743, Korea}

\begin{abstract}
We study a possibility of a strong first order phase transition (FOPT) taking place below the electroweak scale in the context of $U(1)_D$ gauge extension of the standard model. As pointed out recently by the NANOGrav collaboration, gravitational waves from such a phase transition with appropriate strength and nucleation temperature can explain their 12.5 yr data. We first find the parameter space of this minimal model consistent with NANOGrav findings considering only a complex singlet scalar and $U(1)_D$ vector boson. Existence of a singlet fermion charged under $U(1)_D$ can give rise to dark matter in this model, preferably of non-thermal type, while incorporating additional fields can also generate light neutrino masses through typical low scale seesaw mechanisms like radiative or inverse seesaw. 
\end{abstract}

\maketitle

\noindent
{\bf Introduction}: The NANOGrav collaboration has recently released their results for gravitational wave (GW) background produced from a first order phase transition (FOPT) in 45 pulsars from their 12.5 year data \cite{Arzoumanian:2021teu}. According to their analysis, the 12.5 yr data can be explained in terms of a FOPT occurring at a temperature below the electroweak (EW) scale although there exists degeneracy with similar signals generated by supermassive black hole binary (SMBHB) mergers. Last year, the same collaboration also reported evidence for a stochastic GW background with a power law spectrum having frequencies around the nano-Hz regime \cite{Arzoumanian:2020vkk} which led to several interesting new physics explanations; For example, \cite{Blasi:2020mfx, Ellis:2020ena, Bian:2021lmz} studied cosmic string origins and \cite{Ratzinger:2020koh, Addazi:2020zcj, Nakai:2020oit, Bian:2021lmz, Zhou:2021cfu} studied FOPT origin. The pulsar timing arrays (PTAs) like NANOGrav sensitive to GW of extremely low frequencies offer a complementary probe of GW background to future space-based interferometers like eLISA \cite{Caprini:2015zlo,Caprini:2019egz}.

Inspired by the results from NANOGrav explained in terms of a FOPT characterized by the preferred ranges for strength $(\alpha_*)$ as well as temperature $(T_*)$ of the phase transition as shown in \cite{Arzoumanian:2021teu},
we propose a simple model to achieve such a low scale strong FOPT. For our purpose, we introduce a dark $U(1)_D$ gauge symmetry under which only a complex singlet scalar $\Phi$ and a vector like singlet fermion $\Psi$ are charged while all the standard model (SM) particles are neutral. Since the SM particles are neutral under this gauge symmetry, one can evade strong bounds from experiments on the corresponding gauge coupling $g_D$ and gauge boson mass $m_{Z_D}$. We further impose a classical conformal invariance so that $U(1)_D$ symmetry breaking occurs only via radiative effects on scalar potential, naturally leading to a vacuum below EW scale. Then, a strong FOPT can take place at bubble nucleation temperature much below electroweak scale.
For earlier works on FOPT within such Abelian gauge extended scenarios, please refer to \cite{Jinno:2016knw, Mohamadnejad:2019vzg, Kim:2019ogz, Hasegawa:2019amx, Marzo:2018nov, Hashino:2018zsi, Chiang:2017zbz} and references therein.

While such dark phase transition of strongly first order and resulting gravitational waves have been discussed earlier as well, we study this possibility for the first time after NANOGrav collaboration analysed their 12.5 year data in the context of gravitational waves from the FOPT at a low temperature below EW scale \cite{Arzoumanian:2021teu}. In addition, we note that the dark $U(1)_D$ symmetry can also be motivated from tiny neutrino mass and dark matter (DM) which the SM fails to address.
In this work, we examine how tiny neutrino masses can be generated through low scale seesaw mechanism like radiative or inverse seesaw, and a singlet fermion charged under $U(1)_D$ can be a good dark matter candidate while keeping the model parameters consistent with the results from NANOGrav. \\

\noindent
{\bf The Model}: As mentioned above, we consider a $U(1)_D$ extension of the SM. The newly introduced fields are a complex scalar $\Phi$ and a vector like fermion $\Psi$ with $U(1)_D$ charges $2n_1$ and $n_1$, respectively. All the SM fields are neutral under this new gauge symmetry. The zero-temperature scalar potential at tree level is given by
\begin{align}
    V_{\rm tree} = \lambda_H (H^{\dagger} H)^2 + \lambda ( \Phi^{\dagger} \Phi)^2 - \lambda^{'} (\Phi^{\dagger} \Phi) (H^{\dagger} H),
    \label{scalpot1}
\end{align}
where $H$ is the SM Higgs doublet. Note the absence of bare mass squared terms due to the conformal invariance imposed. The vacuum expectation value (VEV) of the singlet scalar, $\langle \Phi \rangle=M/\sqrt{2}$, acquired via the running of the quartic coupling $\lambda$ breaks the gauge symmetry leading to a massive gauge boson $m_{Z_D}=2g_D M$. In order to realize the electroweak vacuum, the coupling $\lambda'$ needs to be suppressed. So in our analysis we neglect the coupling $\lambda'$. We also consider the Yukawa coupling $(y)$ of the scalar singlet with fermion $\Psi$ to be negligible compared to gauge coupling, $g_D \gg y $ for simplicity. This assumption is for simplicity and also to make sure that the SM Higgs VEV does not affect the light singlet scalar mass. Furthermore, the Yukawa coupling $(y)$ is taken to be negligible as to suppress its role in the renormalisation group evolution (RGE) of the singlet scalar quartic coupling.

The total effective potential can be schematically divided into following form:
\begin{align}
V_{\rm tot} = V_{\rm tree} + V_{\rm CW} +V_{\rm th},
\end{align}
where $V_{\rm tree},~V_{\rm CW}$ and $V_{\rm th}$ denote the tree level scalar potential, the one-loop Coleman-Weinberg potential, and the thermal effective potential, respectively. In finite-temperature field theory, the effective potential, $V_{\rm CW}$ and $V_{\rm thermal}$, are calculated by using the standard background field method~\cite{Dolan:1973qd,Quiros:1999jp}. We consider Landau gauge throughout this work. Issues related to gauge dependence in such conformal models have been discussed recently by the authors of \cite{Chiang:2017zbz}. Denoting the singlet scalar as $\Phi=(\phi+M+iA)/\sqrt{2}$, the zero temperature effective potential up to one-loop can be written as \cite{Jinno:2016knw}
\begin{align}
     V_{0} &= V_{\rm tree} + V_{\rm CW}, \nonumber \\
    &= \frac{1}{4} \lambda(t) G^4(t) \phi^4
\end{align}
where $t={\rm log}(\phi/\mu)$ with $\mu$ being the scale of renormalisation. $G(t)$ is given by
\begin{align}
    G(t) = e^{-\int^t_0 dt' \gamma(t')},\; \gamma(t) = -\frac{a_2}{32\pi^2}g^2_D(t),
\end{align}
where we have ignored $\phi$ couplings with $\Psi$ as well as the SM Higgs $H$ 
for simplicity. In the above equation $a_2=24$. The gauge coupling $g_D(t)$ and 
quartic coupling $\lambda(t)$ at renormalisation scale can be calculated by 
solving the corresponding RGE equations. In 
terms of $\alpha_D=g^2_D/4\pi$ and $\alpha_{\lambda}=\lambda/4\pi$, the RGEs are
\begin{align}
    \frac{d\alpha_D(t)}{dt}=\frac{b}{2\pi} \alpha^2_D(t),
\end{align}
\begin{align}
    \frac{d\alpha_{\lambda}(t)}{dt}=\frac{1}{2\pi} \left( a_1 \alpha^2_{\lambda}(t)+8\pi \alpha_{\lambda}(t) \gamma(t)+a_3 \alpha^2_D(t) \right),
\end{align}
where $b=8/3$, $a_1=10$, and $a_3=48$. 
Taking the renormalisation scale $\mu$ to be $M$, the condition $\frac{dV}{d\phi}|_{\phi=M}=0$ leads us to the relation,
\begin{align}
    a_1 \alpha_\lambda(0)^2 + a_3\alpha_{D}(0)^2 + 8\pi \alpha_\lambda(0) &=0,
    \label{eq:eq1}
\end{align}
which makes $\alpha_\lambda(0)$ determined by $\alpha_{D}(0)$.
Since running of the coupling can be solved analytically, the scalar potential can be given by \cite{Jinno:2016knw}
\begin{align}
    V_{0}(\phi,t) &= \frac{\pi \alpha_\lambda(t)}{(1-\frac{b}{2\pi}\alpha_{D}(0)t)^{a_2/b}}\phi^4
\end{align}
where
\begin{align}
    \alpha_{D}(t) &= \frac{\alpha_{D}(0)}{1-\frac{b}{2\pi}\alpha_{D}(0)t} \\
    \alpha_\lambda(t) &= \frac{a_2+b}{2a_1}\alpha_{D}(t) \nonumber \\
    &+ \frac{A}{a_1}\alpha_{D}(t)\tan \left[\frac{A}{b}\ln[\alpha_{D}(t)/\pi] +C\right]  \nonumber \\
    A &\equiv \sqrt{a_1 a_3 - (a_1 + b)^2/4} 
\end{align}
and the coefficient $C$ is determined by Eq. \eqref{eq:eq1}.

Thermal contributions to the effective potential are given by
\begin{align}
V_{\rm th} = \sum_i \left(\frac{n_{\rm B_i}}{2\pi^2}T^4 J_B \left[\frac{m_{\rm B_i}}{T}\right] - \frac{n_{\rm F_{i}}}{2\pi^2}J_F \left[\frac{m_{\rm F_{i}}}{T}\right]\right),
\end{align}
where $n_{B_i}$ and $n_{F_i}$ denote the degrees of freedom (dof) of the bosonic and fermionic particles, respectively.
In this expressions, $J_B$ and $J_F$ functions are defined as follows:
\begin{align}
&J_B(x) =\int^\infty_0 dz z^2 \log\left[1-e^{-\sqrt{z^2+x^2}}\right] \label{eq:J_B},\\
&J_F(x) =   \int^\infty_0 dz z^2 \log\left[1+e^{-\sqrt{z^2+x^2}}\right].
\end{align}
On calculating $V_{\rm th}$, we include a contribution from daisy diagram to improve the perturbative expansion during the phase transition \cite{Fendley:1987ef,Parwani:1991gq,Arnold:1992rz}. The daisy improved effective potential can be calculated by inserting thermal masses into the zero-temperature field dependent masses.
The author of Ref.~\cite{Parwani:1991gq} proposed the thermal resummation prescription in which the thermal corrected field dependent masses are used for the calculation in $V_{\rm CW}$ and $V_{\rm th}$ (Parwani method).
In comparison to this prescription, authors of Ref.~\cite{Arnold:1992rz} proposed alternative prescription for the thermal resummation (Arnold-Espinosa method).
They include the effect of daisy diagram only for Matsubara zero-modes inside $J_B$ function defined above. In our work, we use the Arnold-Espinosa method. As mentioned before, we ignore singlet scalar coupling to fermion and the SM Higgs and hence calculate the field dependent and thermal masses as well as the daisy diagram contribution for vector boson only.

As the evolution has two scales, $\phi$ and $T$, where $T$ is the temperature of the universe, we consider the renormalisation scale parameter $u$ instead of $t$ as 
\begin{align}
    u &\equiv \log(\Lambda/M) \quad {\rm where} \quad \Lambda \equiv {\rm max} (\phi,T)
\end{align}
Note that $\Lambda$ represents the typical scale of the theory. Now, the one-loop level effective potential is given as:
\begin{align}
    V_{\rm tot}(\phi,T) &= V_0(\phi,u) + V_T(\phi,T)   
\end{align}
where 
\begin{align}
    V_T(\phi,T) &= \frac{3}{2}V^B_T(m_V(\phi)/T,T) + V_{\rm daisy}(\phi,T) \\
    V^B_T(x,T) &\equiv \frac{T^4}{\pi^2}\int^\infty_0dz \; z^2\ln[1-e^{-\sqrt{z^2+x^2}}] \nonumber \\
    V_{\rm daisy}(\phi,T) &= -\frac{T}{12\pi}\left[m^3_V(\phi,T) - m^3_V(\phi)\right] \nonumber
\end{align}
wherein, $V^B_T$ is the thermal correction and $V_{\rm daisy}$ is the daisy subtraction \cite{Fendley:1987ef,Parwani:1991gq,Arnold:1992rz}. \\

\begin{figure}
\includegraphics[width=0.45\textwidth]{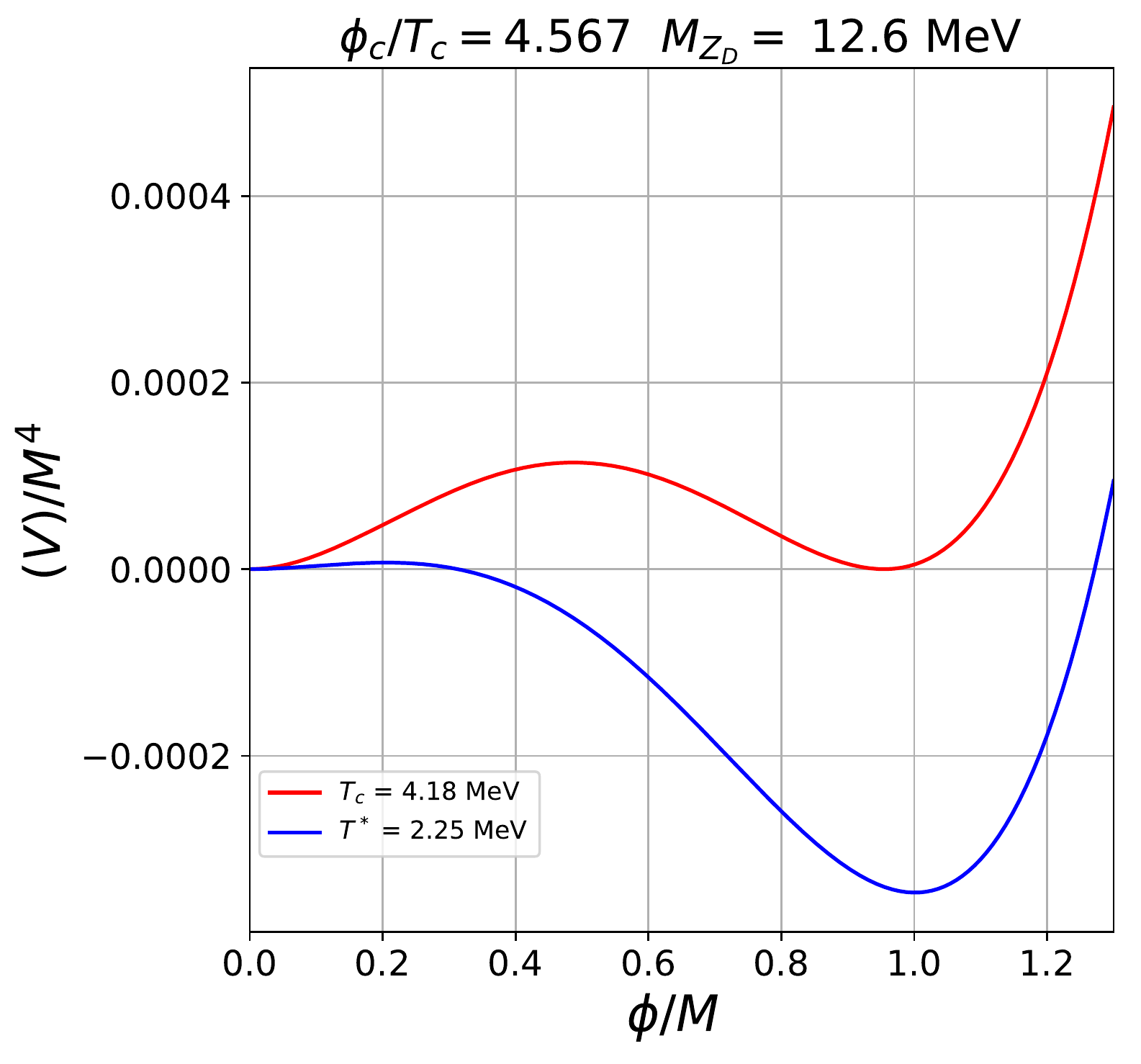}
\caption{Shape of the potential at critical and nucleation temperatures for chosen benchmark $\alpha_*=0.68, T_*=2.25$ MeV, $g_D=0.32, m_{Z_D}=12.6$ MeV.}
\label{fig0}
\end{figure}

\begin{figure*}
\includegraphics[width=0.45\textwidth]{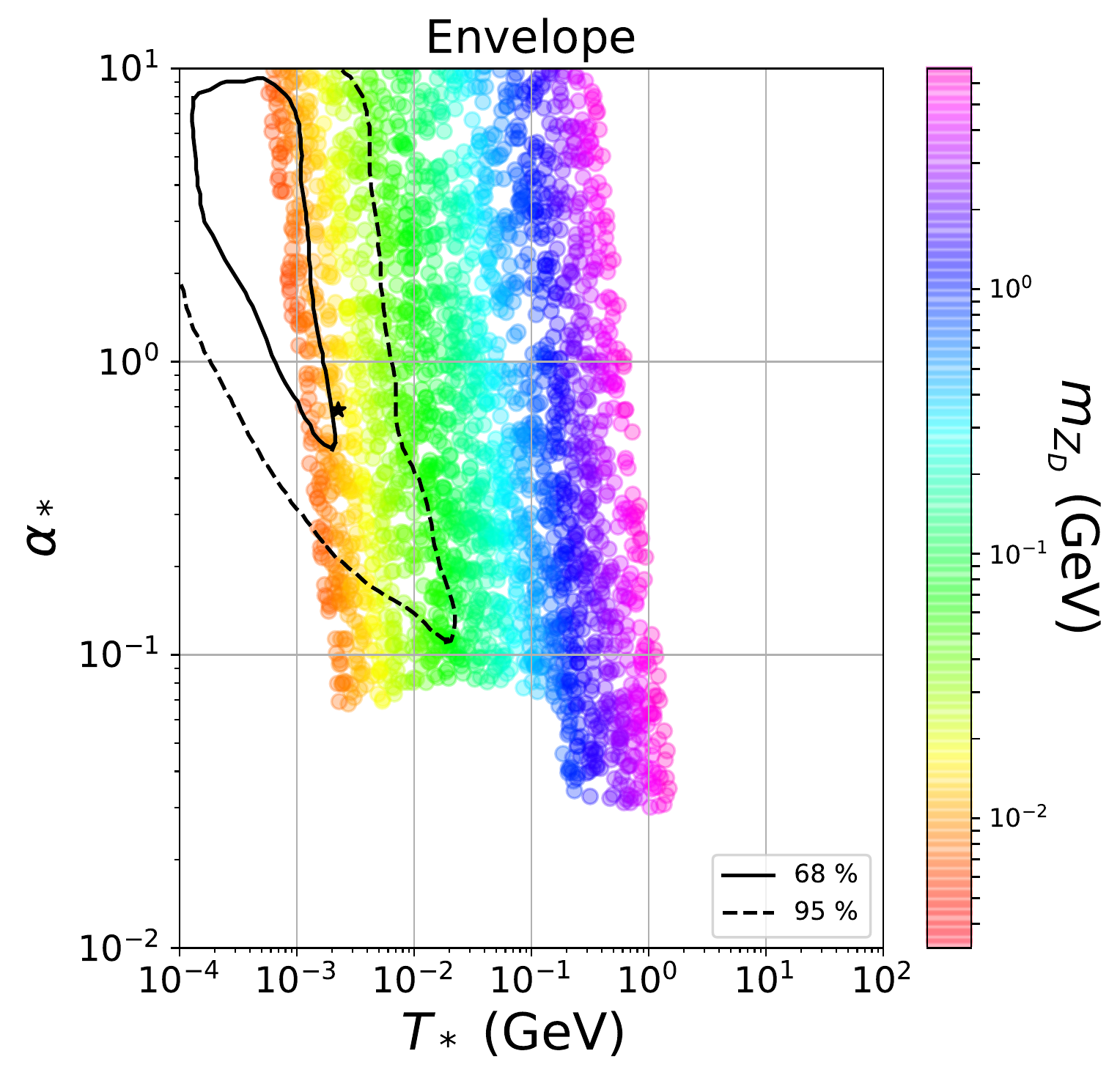}
\includegraphics[width=0.45\textwidth]{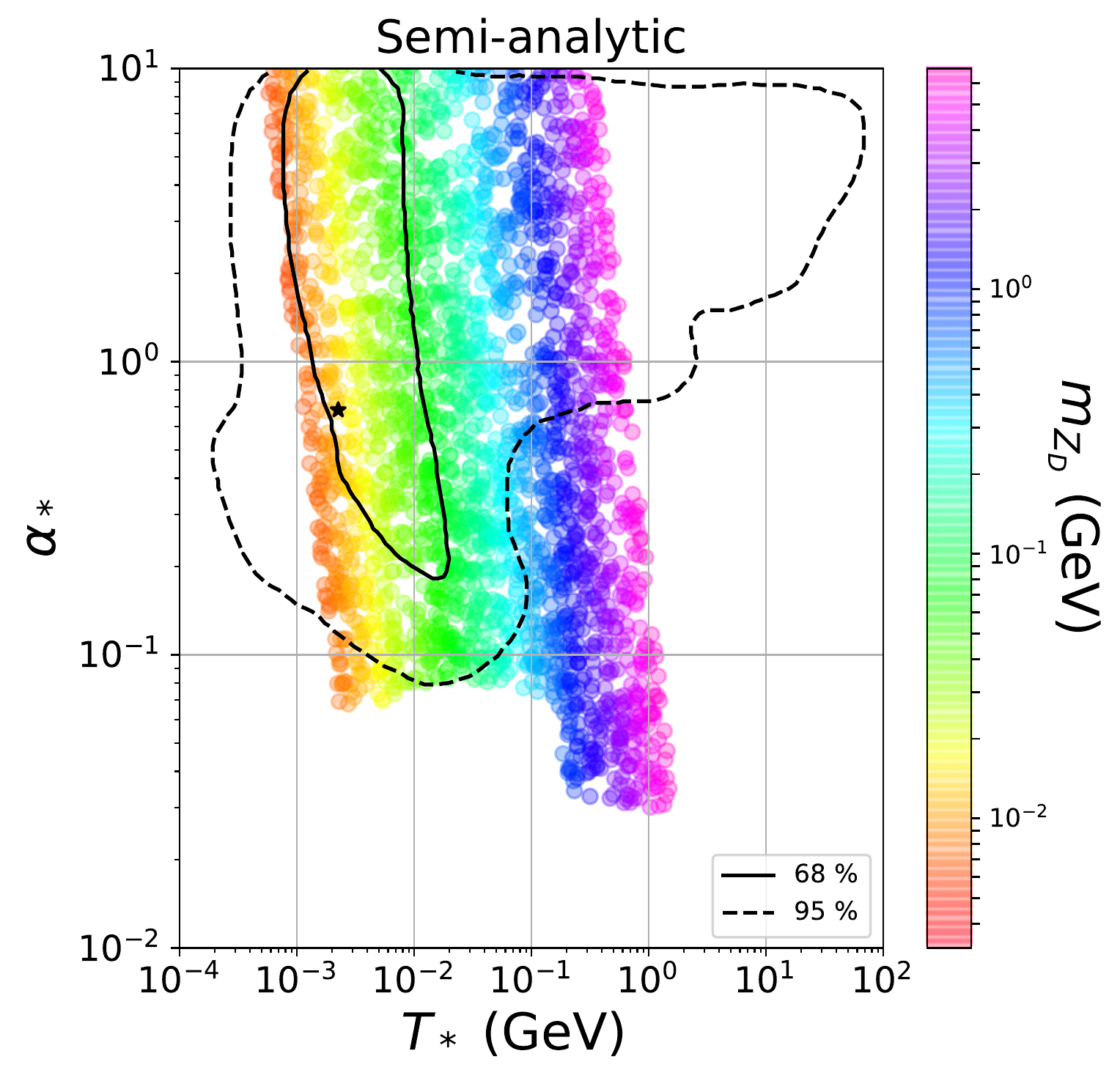}
\includegraphics[width=0.45\textwidth]{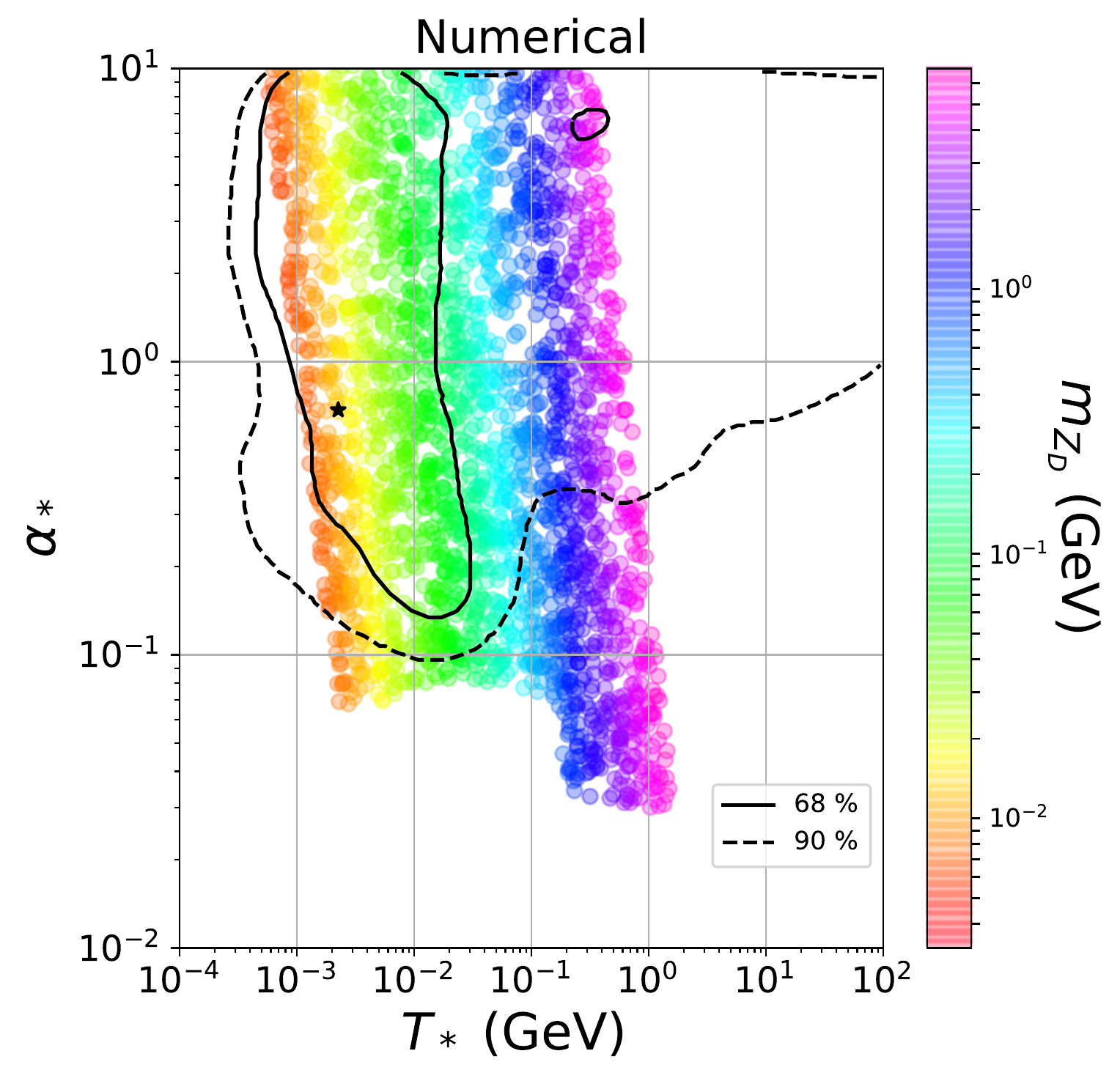}
\caption{Predictions for FOPT parameters in the $\alpha_*-T_*$ plane for our model. The gauge coupling $g_D$ is varied in a range corresponding to $\alpha_D \in 0.002-0.01$. The contours correspond to the confidence levels obtained in\cite{Arzoumanian:2021teu} by using envelope approximation (left panel), semi-analytic approximation (right panel), and numerical results (bottom panel).} 
\label{fig1}
\end{figure*}

\noindent
{\bf First order phase transition}:
The first order phase transitions proceed via tunnelling, and the corresponding spherical symmetric field configurations called bubbles are nucleated followed by expansion and coalescence\footnote{For recent reviews of cosmological phase transitions, refer to \cite{Mazumdar:2018dfl,Hindmarsh:2020hop}.}. The tunnelling rate per unit time per unit volume is given as
\begin{align}
\Gamma (T) = \mathcal{A}(T) e^{-S_3(T)/T},
\end{align}
where $\mathcal{A}(T)\sim T^4$ and $S_3(T)$ are determined by the dimensional analysis and given by the classical configurations, called bounce, respectively. At finite temperature, the  $O(3)$ symmetric bounce solution \cite{Linde:1980tt} is obtained by solving the following equation
\begin{align}
    \frac{d^2 \phi}{dr^2}+\frac{2}{r}\frac{d\phi}{dr} = \frac{\partial V_{\rm tot}}{\partial \phi}\label{eq:bounce diff}.
\end{align}
The boundary conditions for the above differential equation are
\begin{align}
\phi(r\to \infty)= \phi_{\rm false},~~~\left.\frac{d\phi}{dr}\right|_{r=0} =0,\label{eq:boundary condition}
\end{align}
where $\phi_{\rm false}$ denotes the position of the false vacuum. Using $\phi$ governed by the above equation and boundary conditions, the bounce action can be written as
\begin{align}
    S_3 =\int_0^{\infty} dr 4\pi r^2 \left[\frac{1}{2}\left(\frac{d\phi}{dr}\right)^2 +V_{\rm tot}(\phi,T)\right].
    \label{s3eq}
\end{align}
The temperature at which the bubbles are nucleated is called the nucleation temperature $T_*$. This can be calculated by comparing the tunnelling rate to the Hubble expansion rate as
\begin{align}
    \Gamma (T_*) = {\bf H}^4(T_*).
\end{align}
Here, assuming the usual radiation dominated universe, the Hubble parameter is given by ${\bf H}(T)\simeq 1.66\sqrt{g_*}T^2/M_{\rm Pl}$ with $g_*$ being the dof of the radiation component.
Thus, the rate comparison equation above leads to
\begin{align}
    \frac{S_3(T_*)}{T_*} \simeq 140, \label{eq:nucleation temperature}
\end{align}
for $g_*\sim 100$ and $T_* \sim 100$ GeV while for lower temperature near MeV, it comes down to  $g_*\sim 10$. If $\phi(T_*) / T_*>1$ is satisfied, where $\phi(T_*)$ is the singlet scalar VEV at the nucleation temperature, $T=T_*$, the corresponding phase transition is conventionally called \textit{strong} first order. 

By choosing benchmark values as $\alpha_*=0.68, T_*=2.25$ MeV, $g_D=0.32, m_{Z_D}=12.6$ MeV, we can portray the curves of the potential in terms of $\phi/M$ at critical and nucleation temperatures as shown in Fig. \ref{fig0}.
Clearly, we see that $\phi=0$ becomes a false vacuum below the critical temperature $T_c$ and the existence of the barrier at $T_c$ indicates a strong first order phase transition driven by the singlet scalar sector, which triggers bubble production and subsequent production of GW.

The phase transition completes via the percolation of
the growing bubbles. To see when the phase transition finishes, we need to estimate
the percolation temperature $T_p$ at which significant volume of the Universe has been converted from the symmetric to the broken phase.
Following \cite{Ellis:2018mja, Ellis:2020nnr},
$T_p$ is obtained from the probability of finding a point still in the false vacuum
given by
%
\begin{align}
    \mathcal{P}(T) &= e^{-\mathcal{I}(T)} \quad {\rm where}\nonumber \\
    \mathcal{I}(T) &= \frac{4\pi}{3}\int^{T_c}_T \frac{dT'}{T'^4}\frac{\Gamma(T')}{{\bf H}(T')}\left(\int^{T'}_T \frac{d\tilde{T}}{{\bf H}(\tilde{T})}\right)^3.
\end{align}
The percolation temperature is then calculated by using  $\mathcal{I}(T_p) = 0.34$ \cite{Ellis:2018mja} (implying that at least $34\%$ of the comoving volume is occupied by the true vacuum). It is also necessarily required that the physical volume of the false vacuum should be decreased around percolation for successful completion of the phase transition.
This requirement reads
\begin{align}
    \frac{1}{\mathcal{V}_{\rm false}}\frac{d\mathcal{V}_{\rm false}}{dx}&={\bf H}(T)\left(3+T\frac{d\mathcal{I}(T)}{dT}\right)<0. \quad ; x := {\rm time} \label{eq:cond}
\end{align}

 Confirming that this condition is satisfied at the percolation temperature $T_p$, one can ensure that the phase transition successfully completes.
 For the same benchmark values as taken in Fig. \ref{fig0}, we have calculated the percolation temperature $T_p$ and checked that the condition eq.(\ref{eq:cond}) is satisfied.
 The results and values of some parameters are presented in table \ref{tab:GWBP}. \\

\noindent
{\bf Gravitational wave}: As mentioned before, a strong FOPT can lead to the generation of stochastic gravitational wave signals. In particular, GW signals during such a strong FOPT are generated by bubble collisions~\cite{Turner:1990rc,Kosowsky:1991ua,Kosowsky:1992rz,Kosowsky:1992vn,Turner:1992tz}, the sound wave of the plasma~\cite{Hindmarsh:2013xza,Giblin:2014qia,Hindmarsh:2015qta,Hindmarsh:2017gnf} and the turbulence of the plasma~\cite{Kamionkowski:1993fg,Kosowsky:2001xp,Caprini:2006jb,Gogoberidze:2007an,Caprini:2009yp,Niksa:2018ofa}. 

\begin{figure}
\includegraphics[width=0.45\textwidth]{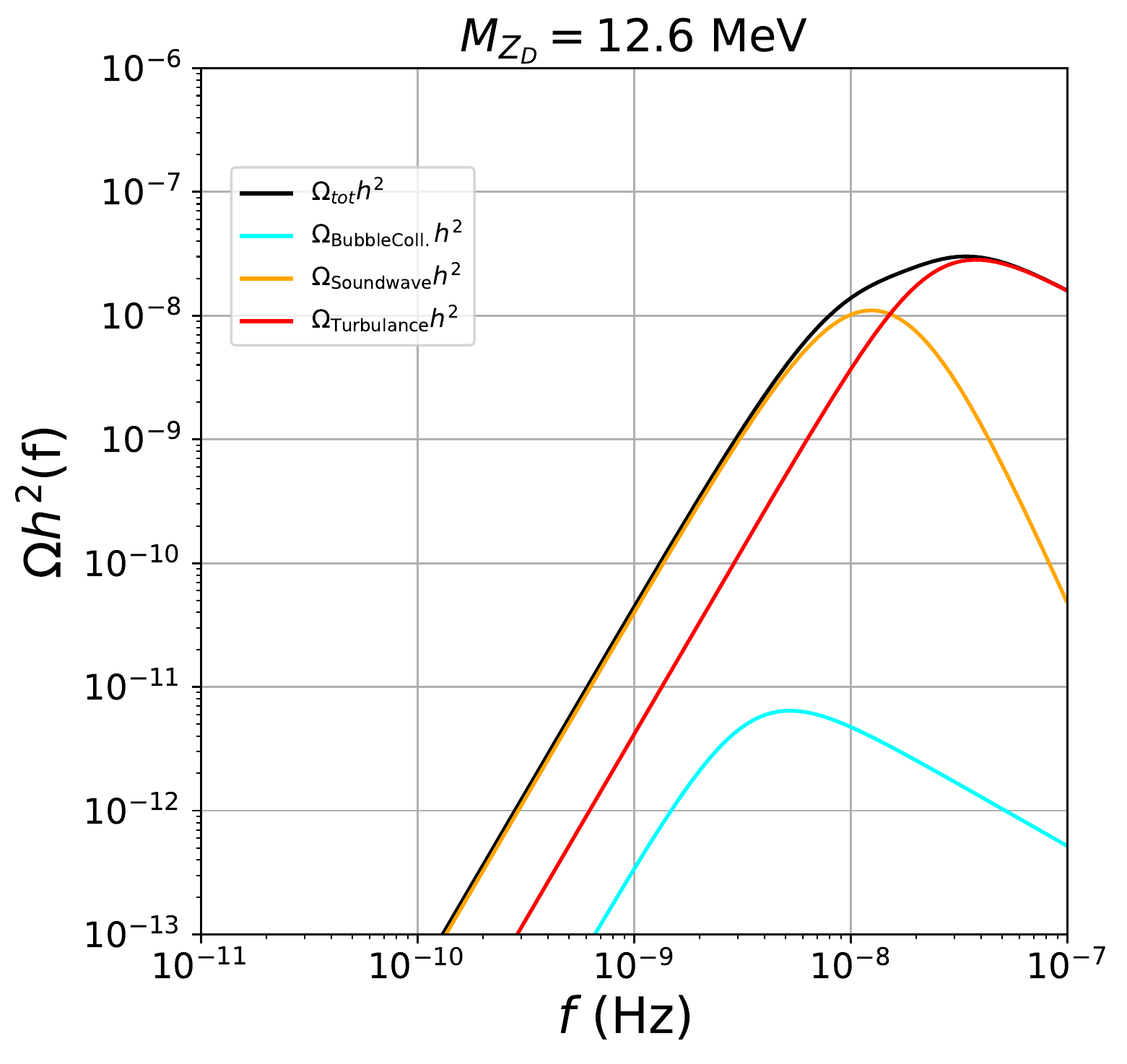}
\caption{GW spectrum $\Omega h^2(f)$ in terms of $f$ for a FOPT with benchmark parameters $\alpha_*=0.68, T_*=2.25$ MeV, $g_D=0.32, m_{Z_D}=12.6$ MeV.
The red, orange, cyan and  black curves correspond to
the individual contribution  from  turbulence of the plasma, sound wave of the plasma, bubble collisions, and the total contribution, respectively.}
\label{fig2}
\end{figure}

The amplitudes of GW depend upon  the ratio of the amount of vacuum energy released by the phase transition to the radiation energy density of the universe, $\rho_{\rm rad}= g_*\pi^2 T^4/30 $, given by
\begin{align}
    \alpha_* =\frac{\epsilon_*}{\rho_{\rm rad}},
\end{align}
with
\begin{align}
    \epsilon_* = \left[\Delta V_{\rm tot} - \frac{T}{4} \frac{\partial \Delta V_{\rm tot}}{\partial T}\right]|_{T=T_*},
\end{align}
where $\Delta V_{\rm tot} \equiv V_{\rm tot}(\phi_{\rm false},T)- V_{\rm tot}(\phi_{\rm true},T)$ is the free energy difference between the false and true vacuum.
$\epsilon_*$ is related to the change in the trace of the energy-momentum tensor across the bubble wall \cite{Caprini:2019egz,Borah:2020wut}. The amplitude of GW is also dictated by the duration of the FOPT, denoted by the parameter $\beta$, defined as \cite{Caprini:2015zlo}
\begin{align}
\frac{\beta}{{\bf H}(T)} \simeq T\frac{d}{dT} \left(\frac{S_3}{T} \right).
\end{align}
Here, $\alpha_*$ and $\beta/{\bf H}(T)$ are evaluated at $T=T_*$. While $S_3$ can be evaluated using eq. \eqref{s3eq}, the effective potential at sufficiently low temperatures i.e $T \ll M$ can be safely approximated as 
\begin{align}
    V_{\rm tot} &\simeq \frac{g^2_{D}(t^{\prime})}{2}T^2\phi^2 + \frac{\lambda_{\rm eff}(t^{\prime})}{4}\phi^4  \label{eq:Veffapp},\\
    \lambda_{\rm eff}(t') &= 4\pi \alpha_\lambda(t')/(1-\frac{b}{2\pi}\alpha_{D}(0)t')^{a_2/b} \nonumber \\
    t^{\prime} &= \ln(T/M).\nonumber 
\end{align}
In such a scenario the action can be approximated to be \cite{Linde:1981zj}
\begin{align}
    S &= \frac{S_3}{T} - 4\ln(T/M) \nonumber \\
    \frac{S_3}{T} &\simeq -9.45 \times \frac{g_{D}(t')}{\lambda_{\rm eff}(t')}
    \label{eq:s3app}
\end{align}
In our estimation for the gravitational wave amplitude we have used the above expressions eq.\eqref{eq:Veffapp} and eq.\eqref{eq:s3app} in calculating $\alpha$, $\beta$ and the percolation temperature $T_p$.

We note that NANOGrav collaboration has 
estimated the required FOPT parameters using thin shell approximation for bubble walls (envelope approximation) \cite{Jinno:2016vai}, semi-analytic approximation \cite{Lewicki:2020azd} as well as full lattice results. Here, we present the predictions of our model against the backdrop of their estimates in Fig. \ref{fig1}.

During a FOPT, there are three sources producing GWs: bubble collisions, sound wave of the plasma, and turbulence of the plasma \cite{Jinno:2016vai, Caprini:2009yp, Hindmarsh:2017gnf, Binetruy:2012ze,Hindmarsh:2015qta,Caprini:2015zlo}. 
These three contributions together give the resultant gravitational wave power spectrum given as \cite{Arzoumanian:2021teu}:
\begin{align}
    \Omega_{\rm GW}(f) &= \Omega_\phi(f) + \Omega_{\rm sw}(f) + \Omega_{\rm turb}(f).
\end{align}
In general, each contribution has its own peak frequency and each GW spectrum can be parametrised in the following way \cite{Arzoumanian:2021teu}
\begin{align}
    h^2\Omega(f) &= \mathcal{R}\Delta(v_w)\left(\frac{\kappa \alpha_*}{1+\alpha_*}\right)^p\left(\frac{{\bf H_*}}{\beta}\right)^*\mathcal{S}(f/f^0_*)
\end{align}
where the pre-factor $\mathcal{R}\simeq 7.69\times 10^{-5}g^{-1/3}_*$ takes in account  the red-shift of the GW energy density, $\mathcal{S}(f/f^0_*)$ parametrises the shape of the spectrum and $\Delta(v_w)$ is the normalization factor which depends on the bubble wall velocity $v_w$. The Hubble parameter at $T=T_*$ is denoted by ${\bf H_*}$. Finally the peak frequency today, $f^0_*$, is related to the value of the peak frequency at the time of emission, $f_*$, as follows:
\begin{align}
    f^0_* &\simeq 1.13\times 10^{-10}{\rm Hz} \left(\frac{f_*}{\beta}\right)\left(\frac{\beta}{{\bf H_*}}\right)\left(\frac{T_*}{ {\rm MeV}}\right)\left(\frac{g_*}{10}\right)^{1/6}
\end{align}
where $g_*$ denotes the number of relativistic degrees of freedom at the time of the phase transition. The values of the peak frequency at the time of emission, the normalisation factor, the spectral shape, and the exponents $p$ and $q$ are given in Table I of \cite{Arzoumanian:2021teu}. The efficiency factors namely, $\kappa_\phi$ is discussed in \cite{Jinno:2016knw, Ellis:2020nnr} and $\kappa_{\rm sw}$ is taken from \cite{Espinosa:2010hh,Borah:2020wut}. On the other hand, the remaining efficiency factor $\kappa_{\rm turb}$ is taken to be approximately $0.1 \times \kappa_{\rm sw}$ \cite{Arzoumanian:2021teu}. The bubble wall velocities are given in \cite{Steinhardt:1981ct, Huber:2013kj,Leitao:2014pda,Dorsch:2018pat,Cline:2020jre}.

\begin{table}[h!]
    \centering
    \begin{tabular}{|c|c|c|c|c|c|}
         \hline 
        $\alpha_*$ &  $(\beta/{\bf H}_*)$ & $T_*$ & $v_w$ & $T_p$ & $ \frac{1}{\mathcal{V}_{\rm false}} \frac{d\mathcal{V}_{\rm false}}{dx}$\\
         \hline
        0.68 & 82.4 & 2.25 MeV & 0.91 & 1.9 MeV & -24.17 GeV\\
         \hline
    \end{tabular}
    \caption{Numerical values of parameters leading to Fig. \ref{fig2}.}
    \label{tab:GWBP}
\end{table}

Based on the formulae presented above and by choosing a benchmark choice of model
as well as FOPT parameters shown in table \ref{tab:GWBP} consistent with NANOGrav data at $95\%$ CL, we calculate 
the individual contributions to GW energy density spectrum $\Omega h^2(f)$ from bubble collisions, sound wave of the plasma, and turbulence of the plasma as well as the total contribution to $\Omega h^2(f)$.
In Fig. \ref{fig2}, 
the red, orange, cyan and  black curves correspond to
the individual contribution  from  turbulence of the plasma, sound wave of the plasma, bubble collisions, and the total contribution to $\Omega h^2(f)$, respectively.
Due to the small value of FOPT strength parameter $\alpha_*$, as anticipated 
from earlier studies \cite{Bodeker:2017cim,Ellis:2019oqb},
the contribution from bubble collision is suppressed as can be seen in Fig. \ref{fig2}. \\

\noindent
{\bf Neutrino mass}: Dark Abelian gauge extension of SM can also be related to the origin of neutrino mass. Neutrino oscillation data suggest tiny but non-vanishing light neutrino masses with two large mixing \cite{Zyla:2020zbs}. Since non-zero neutrino mass and mixing can not be explained in SM, there have been several beyond standard model (BSM) proposals. It turns out that the simplest $U(1)_D$ extension like the one discussed above augmented with additional discrete symmetries of fields can explain the origin of light neutrino mass. Here we briefly mention two such possibilities for neutrino mass origin.

\begin{figure}
\includegraphics[width=0.25\textwidth]{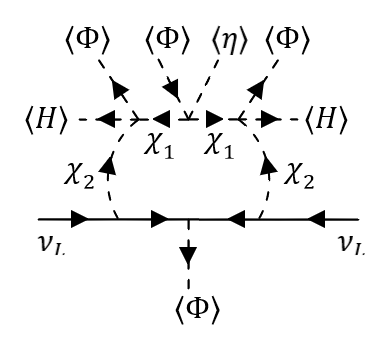}
\caption{One loop origin of light neutrino masses.}
\label{fig3}
\end{figure}

First we discuss a radiative origin of light neutrino masses, a natural origin of low scale seesaw. In addition to the singlet scalar $\Phi$ and the dark fermion $\Psi$ in the minimal model discussed above, we need an additional scalar doublet $\chi$ and a scalar singlet $\eta$ to realise a radiative seesaw. The required field content and their charges under $U(1)_D$ are shown in table \ref{table1}.
\begin{table}
		\begin{tabular}{|c c|c|c|c|c|c|c|}
			\hline
			& &  $\chi_1$ & $\Phi$& $\Psi_{L,R}$& $ \chi_2$ & $\eta$\\ 
			\hline
			& $SU(2)_{L}$  & 1 & 1 & 1& 2 & 1\\
			\hline
			& $U(1)_{D}$  & 1 & 2 & 1& 1 & 0\\
			\hline
		\end{tabular}   
		\caption{New particles for radiative seesaw for neutrino mass with $U(1)_D$ symmetry.}
		\label{table1}
	\end{table} 
The relevant terms of the leptonic Lagrangian are given by
\begin{align}
	-\mathcal{L} & \supset Y_{\nu} \bar{L} \tilde{\chi_2} \Psi_R + Y_L \Phi^{\dagger} \overline{(\Psi_L)^c} \Psi_L
	+ Y_R \Phi^{\dagger} \overline{(\Psi_R)^c} \Psi_R + {\rm h.c.}
	\end{align}
The relevant part of the scalar potential is 
\begin{align}\label{loop-potential}
	V \supset  (\lambda_1 \chi^{\dagger}_2 H \chi^{\dagger}_1 \Phi  + \lambda_2 \chi_1 \chi_1 \Phi^{\dagger} \eta+ {\rm h.c.} )
	\end{align}
The singlet scalar $\eta$, neutral under $U(1)_D$ is introduced in order to avoid terms in the Lagrangian breaking conformal invariance \cite{Ahriche:2016cio}. The $U(1)_D$ symmetry is broken by a nonzero VEV of $\Phi$ to a remnant $Z_2$ symmetry under which $\Psi_{L, R}, \chi_1 , \chi_2$ are odd while all other fields are even. While light neutrino mass can be realised at one loop level with these $Z_2$ odd particles going inside the loop, the lightest $Z_2$ odd particle can be a stable DM candidate. A possible one loop diagram for light neutrino mass is shown in Fig. \ref{fig3}. Since $Z_2$ odd particles take part in the loop, the origin of light neutrino masses is similar to the scotogenic mechanism \cite{Ma:2006km}. The contribution from the diagram shown in Fig. \ref{fig3} can be estimated as
	\begin{equation}
	(m_{\nu})_{ij} \simeq \frac{\lambda^2_1 \langle H \rangle^2 \lambda_2 \langle \Phi \rangle^3 \langle \eta \rangle }{64\sqrt{2}\pi^2} \frac{(Y_{\nu})_{ik} (M_{\Psi})_k (Y^T_{\nu})_{kj}}{M^6_{\chi_2}} 
        I_{\nu} (r_{\chi_1}, r_{k})\,, 
	\end{equation}
 where $(M_{\Psi})_k$ is the mass of pseudo-Dirac fermion states going inside the loop and $I_{\nu}$ is the corresponding loop function written in terms of  $r_{\chi_1} = M^2_{\chi_1}/M^2_{\chi_2}$, and $r_{k} = M^2_{\Psi_k}/M^2_{\chi_2}$ with $M^2_{\chi_1} = (m^2_{\chi_{r1}} + m^2_{\chi_{i1}})/2$ 
and  $M^2_{\chi_2} = (m^2_{\chi_{r2}}+m^2_{\chi_{i2}})/2$. Use of $r, i$ in subscripts denotes real and imaginary neutral parts of the corresponding complex scalar fields. 

We now consider the realisation of another low scale seesaw, namely inverse seesaw with $U(1)_D$ symmetry. It turns out that a minimal $U(1)_D$ gauge symmetry is not sufficient to ensure the required structure of inverse seesaw mass matrix. To have a minimal possibility we consider an additional $Z_4$ discrete symmetry. The new fields and their transformations under the imposed symmetries are shown in table \ref{table2}.
	\begin{table}
		\begin{tabular}{|c c|c|c|c|c|c|c|}
			\hline
			& & $N_R$ & $S_R$  & $\chi $&$\Phi$ & $H_2$ \\ 
			\hline
			& $SU(2)_{L}$ & 1 & 1 & 1 & 1 & 2\\
			\hline
			& $U(1)_{D}$ & 1 & -1 & 0 & 2  & 1\\
			
			\hline
			& $Z_4$ & 1 & i & i & -1  & 1\\
			\hline
		\end{tabular}   
		\caption{New particles for inverse seesaw of neutrino mass with $U(1)_D$ symmetry.}
		\label{table2}
	\end{table} 
The relevant part of the Yukawa Lagrangian is
	\begin{align}\label{neutrino_lagrangian}
	-\mathcal{L} &\supset  Y_\nu \overline{L} \widetilde{H_2} N_R +Y_{NS}N_R S_R \chi^{\dagger} +Y_S S_R S_R \Phi  + {\rm h.c.}
	\end{align}
Clearly, the lepton number violating term involves $\Phi$ which also breaks the $U(1)_D$ symmetry. Therefore, a low scale $U(1)_D$ naturally leads to a tiny lepton number violating term in the inverse seesaw mass matrix. After symmetry breaking, the light neutrino mass is given by
\begin{align}
	m_\nu &\simeq \left (\frac{Y^T_\nu \langle H_2 \rangle}{\sqrt{2}}\right )\frac{1}{M_{NS}}\left (\frac{Y_S \langle \Phi \rangle}{\sqrt{2}}\right )\frac{1}{M_{NS}} \left (\frac{Y_\nu \langle H_2 \rangle}{\sqrt{2}} \right )
	\end{align}
where $M_{NS}= \frac{Y_{_{NS}} \langle \chi \rangle}{\sqrt{2}}$. 

Thus, in both the examples discussed here, the low scale $U(1)_D$ symmetry can play non-trivial role in light neutrino mass generation even though all the SM fields are neutral under this symmetry. \\
\begin{figure*}
\includegraphics[width=0.45\textwidth]{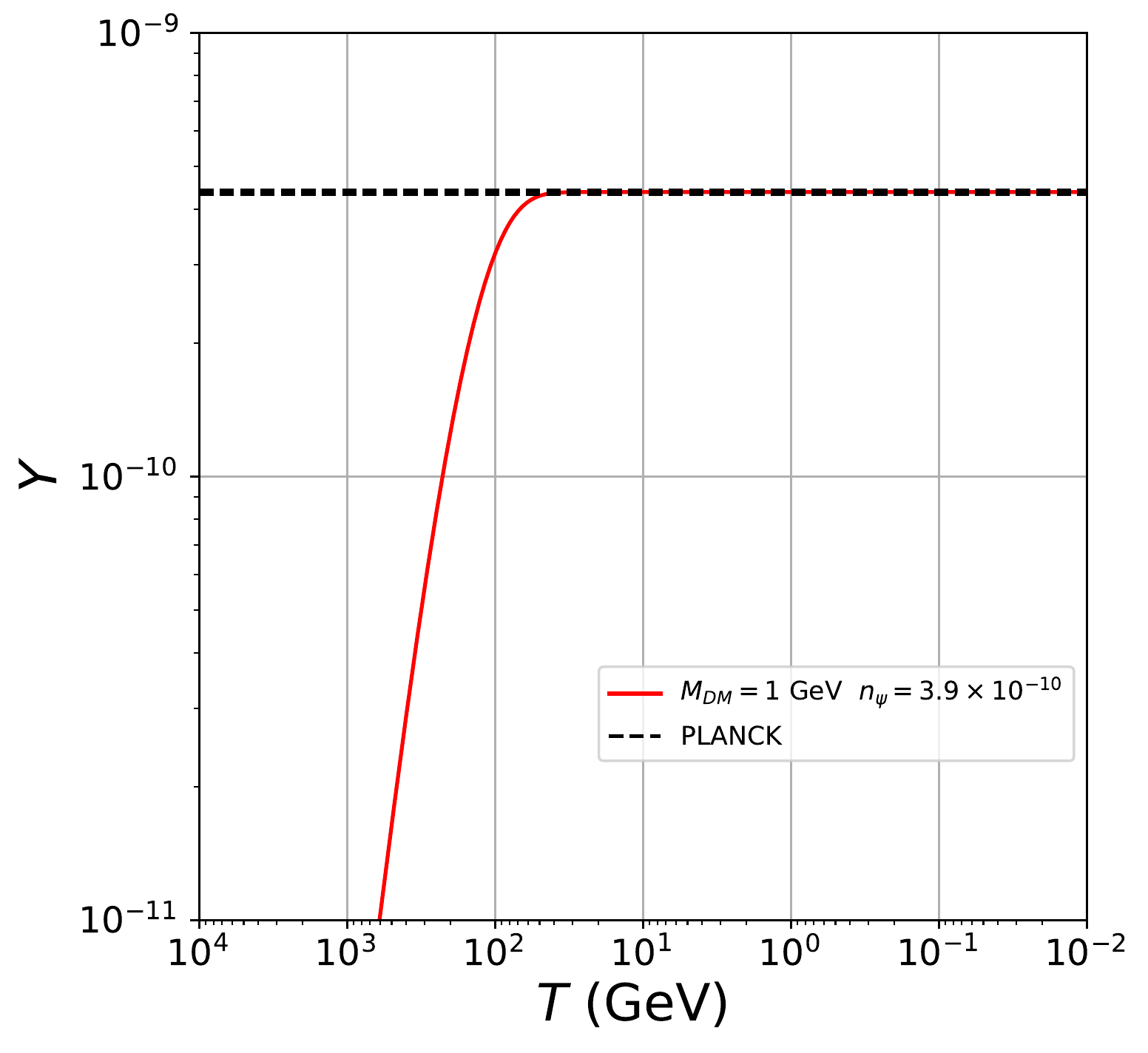}
\includegraphics[width=0.45\textwidth]{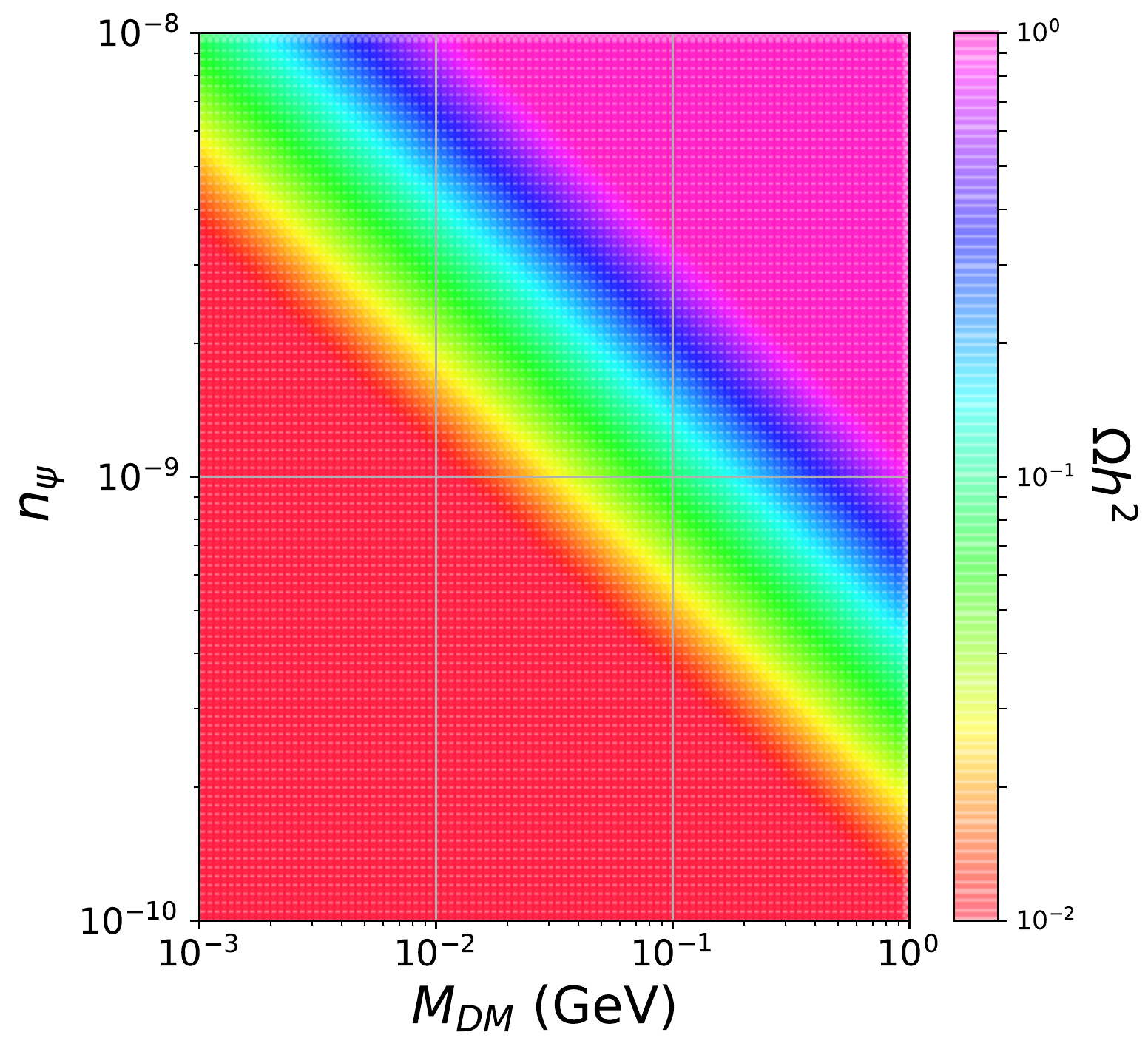}
\caption{Left panel: Comoving DM number density $Y$ {\it vs.} $T$ for DM mass $M_{\rm DM}=1$ GeV and its $U(1)_D$ charge $n_\psi = 3.9 \times 10^{-10}$. Right panel: The regions of $M_{\rm DM}-n_{\psi}$ parameter space giving rise to correct DM relic for $g_D=0.32, m_{Z_D}=12.6$ MeV.} 
\label{fig5}
\end{figure*}

\noindent
{\bf Dark matter and cosmological constraints}:
Evidences from astrophysics and cosmology suggest the presence of a non-baryonic form of matter giving rise to approximately $26\%$ of the present universe's energy density \cite{Zyla:2020zbs}. The simplest possibility is to consider a vector like fermion $\Psi$ having charge $n_{\psi}$ under $U(1)_D$. Depending on the strength of gauge interactions, the relic abundance of DM can be realised either via thermal or non-thermal mechanisms. While the $U(1)_D$ gauge coupling was kept large in the analysis for FOPT and GW above, DM interactions with the SM can still be suppressed due to small kinetic mixing between $U(1)_D$ and $U(1)_Y$. However, in the discussion on neutrino mass, we have introduced additional fields charged under both SM and $U(1)_D$ gauge symmetries. This will keep the one loop kinetic mixing between $U(1)_D$ and $U(1)_Y$ suppressed but still large enough to produce $Z_D$ in equilibrium. Thus, a light gauge boson with not too small kinetic mixing with $U(1)_Y$ can decay into SM leptons at late epochs (compared to neutrino decoupling temperature $T^{\nu}_{\rm dec} \sim \mathcal{O}(\rm MeV)$ increasing the effective relativistic degrees of freedom which is tightly constrained by Planck 2018 data as ${\rm N_{eff}= 2.99^{+0.34}_{-0.33}}$ \cite{Aghanim:2018eyx}. Such constraints can be satisfied if $m_{Z_D} \gtrsim \mathcal{O}(10 \; {\rm MeV})$ \cite{Ibe:2019gpv, Escudero:2019gzq} which agrees with the benchmark value chosen in our FOPT and GW analysis. On the other hand, taking $m_{Z_D}$ to much higher regime will not explain the NANOGrav data. Therefore, we keep its benchmark at minimum allowed value. Similar bound also exists for thermal DM masses in this regime which can annihilate into leptons. As shown by the authors of \cite{Sabti:2019mhn}, such constraints from the big bang nucleosynthesis (BBN) as well as the cosmic microwave background (CMB) measurements can be satisfied if $M_{\rm DM} \gtrsim \mathcal{O}(1 \; {\rm MeV})$. On the other hand, constraints from CMB measurements disfavour such light sub-GeV thermal DM production in the early universe through s-channel annihilations into SM fermions \cite{Aghanim:2018eyx}. Since fermion singlet DM in our model primarily annihilates via s-channel annihilations mediated by $Z_D$ only, cosmological constraints are severe for thermal DM mass around or below 10 MeV.

Due to the tight cosmological constraints on thermal DM with mass below 10 MeV as discussed above, we consider a non-thermal DM scenario, also known as the feebly interacting massive particle (FIMP) paradigm \cite{Hall:2009bx}. While we can not make $g_D$ very small, in order to satisfy the FOPT and GW criteria, we choose $U(1)_D$ charge of DM $n_{\psi}$ to be very small\footnote{FIMP DM in similar Abelian gauge model with tiny $U(1)$ charge of DM was studied in earlier works like, for example, \cite{Biswas:2016yjr} where authors studied $L_{\mu}-L_{\tau}$ gauge symmetry.}.
For DM mass above $Z_D$, it can be produced in the early universe via annihilation of SM bath particles into DM, mediated by $Z_D$. On the left panel of Fig. \ref{fig5}, we show the evolution of comoving DM number density $Y$ for DM mass $M_{\rm DM}=1$ GeV and its $U(1)_D$ charge $n_\psi = 3.9 \times 10^{-10}$. The kinetic mixing of $Z_D$ with $U(1)_Y$ of the SM is taken to be approximately $\epsilon \sim g_D g'/(16 \pi^2)$, similar to one-loop mixing. Clearly, DM with negligible initial abundance freezes in and gets saturated at lower temperatures, giving rise to the required relic density. On the right panel of Fig. \ref{fig5}, we show the parameter space in terms of $M_{\rm DM}-n_{\psi}$ giving rise to correct DM relic while keeping $U(1)_D$ sector parameters fixed at $g_D=0.32$ and $m_{Z_D}=12.6$ MeV. Since DM mass is varied all the way upto 1 MeV for the right panel plot of Fig. \ref{fig5}, which is below the $Z_D$ mass threshold, we consider both annihilation and decay contributions to DM relic. Clearly, smaller values of $n_{\psi}$ requires larger DM mass to satisfy the relic criteria. This is because, smaller DM coupling leads to smaller non-thermal abundance and hence larger mass is required to generate the observed relic abundance. While we skip other phenomenological signatures of such DM, such sub-GeV DM can have very interesting phenomenology in the context of latest experiments like XENON1T \cite{Aprile:2020tmw} as has been discussed by \cite{Borah:2020smw,Borah:2020jzi,Dutta:2021wbn,Borah:2021jzu} among others. Such Dirac fermion DM, upon receiving a tiny Majorana mass contribution from singlet scalar, as discussed in the context of radiative neutrino mass above can give rise to inelastic DM \cite{TuckerSmith:2001hy, Cui:2009xq} with interesting DM phenomenology \cite{Song:2021yar}. \\

\noindent
{\bf Conclusion}: Motivated by the recent NANOGrav collaboration's analysis of their 12.5 yr data implying a possible origin of stochastic GW spectrum from a first order phase transition below EW scale, we revisit the simplest possibility of a dark Abelian gauge extension of the SM. While the SM fields are neutral under this gauge symmetry, a complex scalar singlet with non-vanishing gauge charge can lead to the necessary symmetry breaking. We further consider a classical conformal invariance such that the symmetry breaking occurs through radiative corrections to the scalar potential, keeping the model minimal. While additional dark fermions can be introduced in order to explain the origin of dark matter, for the phase transition details we confine ourselves to only the singlet scalar - vector boson interactions, ignoring other scalar portal or Yukawa interactions for simplicity. We perform a numerical scan to show how a light  gauge boson $Z_D$ in sub-GeV scale can explain the FOPT parameters given by \cite{Arzoumanian:2021teu} in order explain their data. We have also commented on the possibility of connecting such $U(1)_D$ models to light neutrino mass and dark matter in a common setup. Due to tight cosmological constraints on such light vector boson $Z_D$ as well as DM whose interactions with the SM sector are mediated by $Z_D$ via kinetic mixing, we consider a non-thermal DM scenario. By choosing $U(1)_D$ sector parameters in a way which satisfies NANOGrav data, we perform a numerical scan over DM mass in sub-GeV range and its $U(1)_D$ charge which can give rise to the correct non-thermal DM relic.

Due to complementary nature of observable signatures of such minimal models, specially in the context of GW from FOPT as well as typical sub-GeV dark matter signatures, near future experiments should be able to do more scrutiny of such predictive scenarios. Additionally, while PTAs like NANOGrav offer a complementary GW window to proposed space-based interferometers, more data are need to confirm whether this is a clear detection of GW and whether it is due to FOPT or astrophysical sources like SMBHB mergers (Possible ways of distinguishing cosmological backgrounds from astrophysical foregrounds have been discussed recently in \cite{Moore:2021ibq}).

\noindent
\acknowledgements
DB acknowledges the support from Early Career Research Award from DST-SERB, Government of India (reference number: ECR/2017/001873). AD and SKK are supported in part by the National Research Foundation (NRF) grants NRF-2019R1A2C1088953.


\end{document}